\documentstyle[12pt]{article}
\advance\voffset by -1.0in
\advance\hoffset by -0.75in
\input amssym.def
\input amssym
\newcommand{\Hil}{{\cal H}}
\newcommand{\kay}{{\cal K}}
\newcommand{\ze}{{\Bbb Z}}
\newcommand{\re}{{\Bbb R}}
\newcommand{\enn}{{\Bbb N}}
\newcommand{\reg}[1]{(\ref{#1})}
\begin{document}
\title{On the Uniqueness of the Twisted Representation in the
$Z\!\!\!Z_2$ Orbifold Construction of a Conformal Field Theory from a
Lattice}
\author{P.S. Montague\\
Department of Physics and
Mathematical Physics\\
University of Adelaide\\
Adelaide\\
South Australia 5005\\
Australia}
\maketitle
\begin{abstract}
Following on from recent work describing the representation content
of a meromorphic bosonic conformal field theory in terms of a certain
state inside the theory corresponding to a fixed state in the
representation, and using work of Zhu on a correspondence between the
representations of the conformal field theory and representations of a
particular associative algebra constructed from it, we construct a
general solution for the state defining the representation and identify
the further restrictions on it necessary for it to correspond to a ground
state in the representation space. We then use this general theory to analyze
the representations of the Heisenberg algebra and its $\ze_2$-projection.
The conjectured uniqueness of the twisted representation is shown explicitly,
and we extend our considerations to the reflection-twisted FKS construction
of a conformal field theory from a lattice.
\end{abstract}
\section{Introduction}
In \cite{PSMreps} an argument was given for what we term the uniqueness
of the twisted representation of the reflection-twisted
projection of an FKS lattice conformal field theory.
In this paper, we shall present an alternative and more explicit proof,
as well as introducing a method which is of more general applicability
and interest in its own right.

Our original motivation for proving the above-mentioned uniqueness
was to enable the completion of the argument of \cite{DGMtriality,DGMtrialsumm}
which extended the ``triality'' of Frenkel, Lepowsky and Meurman
\cite{FLMbook} to a
more general class of theories than just the Moonshine module for the Monster.
While this is of sufficient import, further motivation (beyond the obvious
intrinsic interest of the problem) exists. For example,
this work and its generalization to higher order twists is clearly
of relevance to orbifold constructions of conformal field theories in which
there has been much interest as providing constructions of more physically
realistic string models
\cite{DFMS,stringorb1,stringorb2,CorrHoll1,CorrHoll2,Hollthesis,Vafa}.
It is also
hoped that these and similar uniqueness arguments may be used to help
complete the classification of the self-dual theories at $c=24$
\cite{PSMorb}, {\em i.e.}
in verifying that any conformal field theory corresponding to one of the
algebras listed by Schellekens in \cite{Schell:Venkov,SchellComplete} is
unique and further it is hoped that the abstract
techniques developed here for finding
representations may help in the construction of orbifolds corresponding to
these algebras in the first place. In addition, this viewpoint may provide
a deeper understanding of what precisely is meant by the dual of a
conformal field theory and the concept of self-duality \cite{PGmer}.

It has long been suspected that the known meromorphic
representations for the
reflection-twisted \cite{DGMtwisted,Lepowsky} projection of an FKS (or
``untwisted'' \cite{FK,S})
lattice conformal field theory are complete. We provide details later
in the paper of the precise structure of these objects.
We merely note here that the known representations comprise
those trivially inherited from the unprojected untwisted lattice theory
together with essentially ({\em i.e.} up to inequivalent
ground states) two ``twisted''
representations. Since the twisted representations
are both projections of a single non-meromorphic
representation of the unprojected untwisted theory, we refer to this
conjecture as the ``uniqueness
of the twisted representation''.

Dong \cite{dongmoonrep} has proven the result in the specific case in which
the underlying lattice is taken to be the Leech lattice. This argument appears
difficult to generalize however.

In \cite{PSMreps} we found a description of a representation of
a conformal field theory $\Hil$ in terms of some state $P$ in the theory
corresponding
to a particular (fixed) state in the representation space. We then in some
sense inverted the argument to construct a representation from $P$ of a
larger conformal field theory in which $\Hil$ is embedded as a sub-conformal
field theory \cite{thesis}. This allowed us to extend a representation of
the $\ze_2$-projected theory to one of the original unprojected FKS lattice
conformal field theory. (Similar ideas appear independently in
\cite{DongInduced}.)
The representations of this are well known
\cite{dongtwisrep} and the required uniqueness follows. However, several
crucial technical points are ignored. For example it is not clear
that the matrix elements of the representation defined in terms of $P$
(see section \ref{regurge} for more details)
have the required analytic properties for the larger theory, or indeed
that the Hilbert space for the representation induced by these matrix elements
is even a Hilbert space.
For these reasons we seek a more direct proof
of the uniqueness of the twisted representation.
We can thus really, in the following, only treat the conditions
on $P$ derived in \cite{PSMreps} as necessary and not sufficient for
the existence of a representation. Even in the absence of such problems
with the extension (as in the ``induced modules'' of \cite{DongInduced}),
the following explicit analysis of the representations of the $\ze_2$-projected
theory is clearly of general applicability and interest in its own right,
and is a step towards a full understanding of the origin and nature
of the twisted structure.

The layout of the paper is as follows. In section \ref{regurge} we summarize
the results of \cite{PSMreps} on the description of a representation of
a conformal field theory in terms of some state in the theory corresponding
to a particular (fixed) state in the representation space. Then in section
\ref{develop} we develop these results further, particularly in the light
of work of Zhu \cite{Zhu}, and produce a general solution to the equations
in \cite{PSMreps} as well as necessary and sufficient conditions for the
state to correspond to a ground state in the representation space ({\em i.e.}
reducing the vast degeneracy of solutions -- which we must do
if we are to have any hope of
using this technique to classify possible representations). In section
\ref{simple} we look at a simple application, namely the one-dimensional
Heisenberg algebra, before attempting an analysis of the
representations of the $\ze_2$-projection
of the Heisenberg algebra (projection onto an even number of oscillators)
in section \ref{notsosimple} and then an analysis considering also
the existence of non-zero momentum eigenstates.
This is finally extended to a consideration of
the reflection-twisted projection of the FKS theory.
We end in section \ref{conclusions}
with some general comments and some speculations about applications
to higher order twisted modules and the general construction of orbifolds
>from conformal field theories.
\section{Representations of Conformal Field Theories}
\label{regurge}
We summarize here the main results of \cite{PSMreps}.

Let us first establish our notation. We define a conformal field theory
(strictly a bosonic, hermitian, meromorphic conformal field theory)
to consist of a Hilbert space $\Hil$, two fixed states $|0\rangle$, $\psi_L\in
\Hil$,
 and a set $\cal V$ of ``vertex operators'',
{\em i.e.} linear operators
$V(\psi,z):\Hil\rightarrow\Hil$, $\psi\in\Hil$ parameterized by a complex
parameter $z$ such that $V(\psi_1,z_1)V(\psi_2,z_2)\cdots$ makes
sense for $|z_1|>|z_2|>\cdots$,
\begin{equation}
V(\psi,z)|0\rangle=e^{zL_{-1}}\psi\,,
\end{equation}
\begin{equation}
\label{locality}
V(\psi,z)V(\phi,w)=V(\phi,w)V(\psi,z)
\end{equation}
(in the sense that appropriate analytic continuations of matrix
elements of either side agree) and
\begin{equation}
V(\psi_L,z)\equiv\sum_{n\in\ze}L_nz^{-n-2},,
\end{equation}
where
\begin{equation}
[L_m,L_n]=(m-n)L_{m+n}+{c\over 12}m(m^2-1)\delta_{m,-n}\,,
\end{equation}
(the constant $c$ is the ``central charge'' of the theory).
See \cite{DGMtriality} for a full discussion of this definition (as well
as the more technical axioms omitted here).

We define a representation of this theory to consist of a Hilbert space
$\kay$ and a set of linear (vertex) operators $U(\psi,z):\kay\rightarrow
\kay$, $\psi\in\Hil$, such that
\begin{equation}
\label{duality}
U(\psi,z)U(\phi,w)=U(V(\psi,z-w)\phi,w)\,,
\end{equation}
and also $U(|0\rangle)\equiv 1$. Note that a relation identical to
\reg{duality} is satisfied by the $V$'s as a consequence of the above axioms
\cite{PGmer,DGMtriality}. This representation is said to be meromorphic if
matrix elements of the $U$'s are meromorphic functions of the complex
arguments.

In \cite{PSMreps} we showed that, given a fixed quasi-primary
state $\chi$ in
the representation, there is some state $P(z)$ in
$\Hil$ such that
\begin{eqnarray}
\langle\chi|U(\psi_1,z_1) U(\psi_2,z_2)&\ldots&
U(\psi_n,z_n)|\chi\rangle=\nonumber\\
\langle P(z_n^\ast)|V(\psi_1,z_1-z_n) V(\psi_2,z_2-z_n)&\ldots&
V(\psi_{n-1},z_{n-1}-z_n)|\psi_n\rangle\,,
\label{maindef}
\end{eqnarray}
[Note that the definition of $P$ given in \cite{PSMreps} rather assumes
an orbifold-like structure, and instead we might define $P$ by
\begin{equation}
\langle P(z^\ast)|\psi\rangle=\langle\chi|U(\psi,z)|\chi\rangle\,,
\end{equation}
for all $\psi\in\Hil$. This is essentially just taking the projection
onto $\Hil$ of the definition in \cite{PSMreps}. We shall see examples
of this distinction later.]
We derived necessary (and we believe also
sufficient) conditions on $P(z)\equiv\sum_{n\in\ze_+}P_nz^{-n}$,
$P_n$ of conformal weight $n$, in order that
these matrix elements
be the matrix elements of a representation.
These are (for a {\em real} representation \cite{PSMreps,DGMtriality})
\begin{eqnarray}
\langle 0|P(z)\rangle&=&1\nonumber\\
\overline{P(z)}&=&P(-z^\ast)\nonumber\\
e^{(z-w)L_1}P(w)&=&P(z)\nonumber\\
e^{zL_{-1}}P(-z)&=&P(z)\,,
\label{constraints}
\end{eqnarray}
where $\psi\mapsto\overline\psi$ is an antilinear map on $\Hil$ corresponding
to hermitian conjugation \cite{DGMtriality}.
\section{Solutions for $|P\rangle$ Corresponding to a Ground State}
\label{develop}
Note that the previous section involved an arbitrary choice of state
$\chi$ (in fact $\chi$ is constrained to be of unit norm and real, {\em i.e.} $
\overline\chi=\chi$)
in the Hilbert space for the representation. Thus any solution
to the equations \reg{constraints} will reflect this, and we expect
an infinite number of solutions. Clearly this is of no use if we wish to
use this to try to restrict the number of possible representations.
In this section, we impose the further condition that $\chi$ lies
in the ground state of the representation module.

The work of Zhu \cite{Zhu} appears to pursue many of the same ideas
put forward in \cite{PSMreps}. In particular, Zhu develops a 1-1
correspondence between representations of the conformal field theory
$\Hil$ and representations of an associative algebra which he terms
$A(\Hil)$ (which thus in some way corresponds to the object $P(z)$ above,
though at the moment the exact correspondence is unclear). In the
course of this, he defines a bilinear operation $*$  on $\Hil$ and a
two-sided ideal for $*$ denoted $O(\Hil)$
by (rewriting his definitions in terms of the explicit modes of the
vertex operators)
\begin{eqnarray}
\label{stardef}
\psi*\phi&=&\sum_{r=0}^h\left({h\atop r}\right)V(\psi)_{-r}\phi\nonumber\\
O(\Hil)&=&{\rm span}\ \{O(\psi,\phi)\}\nonumber\\
O(\psi,\phi)&=&\sum_{r=0}^h\left({h\atop r}\right)V(\psi)_{-r-1}\phi\,,
\end{eqnarray}
where $\psi$ is of conformal weight ($L_0$ eigenvalue) $h$.
Further, he shows that for $\psi\in O(\Hil)$ and $\chi$ in the ground state
of a representation of $\Hil$ with vertex operators $U$, $U(\psi)_0\chi=0$.
Also we have the useful result
\begin{equation}
\label{product}
U(a*b)_0\chi=U(a)_0U(b)_0\chi\,.
\end{equation}

We now apply these results in the context of our notation.
\subsection{Restriction of $\chi$ to a ground state}
Define $P=P(1)$ ({\em i.e.} $P(z)=z^{-L_0}P$). Then note from
\reg{maindef} that
\begin{equation}
\label{zeromode}
\langle\chi|U(\psi)_0|\chi\rangle=\langle P|\psi\rangle\,.
\end{equation}
Hence, for $\psi\in O(\Hil)$,
\begin{equation}
\label{whodoyoudo}
\langle P|\psi\rangle=0\,.
\end{equation}
This condition is also sufficient for the state $\chi$ to which $P$
corresponds to lie in the ground state. To see this, we first note from
Zhu that
\begin{equation}
\label{howdy}
O(a,b)=c(a)*b\,,
\end{equation}
where
\begin{equation}
\label{howdydoody}
c(a)\equiv(L_{-1}+L_0)a\,.
\end{equation}
Following a calculation of Zhu (pages 15-16 of \cite{Zhu}), we see that
\begin{equation}
\langle\chi|U(c(a)*b)|\chi\rangle=
\langle\chi|\sum_{i\in{\enn}}\left(
U_{-i}(c(a))U_i(b)+U_{-i-1}(b)U_{i+1}(c(a))\right)|\chi\rangle=0\,.
\end{equation}
But we trivially have
\begin{equation}
\label{triv}
U_i(L_{-1}a)=-(i+{\rm wt}\,a)U_i(a)\,,
\end{equation}
and thus
we find that orthogonality of $P$ to $O(\Hil)$ is equivalent to
\begin{equation}
\langle\chi|\sum_{i\in{\enn}}i\left[U_{-i}(a)U_i(b)-
U_{-i}(b)U_i(a)\right]|\chi\rangle=0\,,
\end{equation}
for all states $a$ and $b$ (extending $a$ to an arbitrary state
by linearity).

In particular, taking $b=c(\overline a)$ for $a$ quasi-primary
and using \reg{triv} again, we get
a sum of norms, and hence deduce that $U_i(a)|\chi\rangle=0$ for
all quasi-primary $a$ and all $i>0$. The statement for all states
follows simply from \reg{triv}, and thus we see that $\chi$ is a
highest weight state, as required.

Thus, our goal now is to solve the equations \reg{constraints}
for $P$ and impose orthogonality to $O(\Hil)$. This should now provide
a finite set of solutions. Because of our reluctance to conclude that these
conditions are also sufficient, we cannot say that each solution corresponds
to a representation, but if we can find a representation corresponding to
each possible solution then the classification of the possible
representations will be complete.
\subsection{General solution for $P$}
We see that the first two equations in \reg{constraints} are
satisfied trivially by taking
\begin{equation}
P=|0\rangle+\sum_{\psi \rm{real}}\alpha_\psi\psi\,,
\end{equation}
where the sum is taken over all real states
(of strictly positive conformal weight) and the coefficients $\alpha_\psi$
are real multiples of $i^{{\rm wt}\,\psi}$.

The third equation we see is equivalent to
\begin{equation}
\label{L1}
(L_1+L_0)P=0\,.
\end{equation}
The most general solution to this is to write
\begin{equation}
P=|0\rangle+\sum_{i,n}\beta_{i,n}{L_{-1}}^n\phi_i\,,
\end{equation}
where the $\phi_i$ are the real
quasi-primary states in $\Hil$ and $\beta_{i,n}$ satisfy
\begin{equation}
\beta_{i,n+1}=-{1\over 2}{(n+h_i)\beta_{i,n}\over (n+1)h_i+{1\over 2}n(n+1)}\,,
\end{equation}
where $h_i$ is the conformal weight of $\phi_i$.
The solution to this is
\begin{equation}
\label{beta}
\beta_{i,n}=\beta(h_i,n)\beta_i\,,
\end{equation}
for arbitrary $\beta_i$, where
\begin{equation}
\beta(h,n)={(-1)^n(2h-1)!(h+n-1)!\over (h-1)!n!(2h+n-1)!}\,.
\end{equation}

In order to satisfy the final relation of \reg{constraints},
we are required to show that $e^{{1\over 2}x}F_h(x)$ is an even function,
where
\begin{equation}
F_h(x)=\sum_{n=0}^\infty \beta(h,n)x^{n+h}\,.
\end{equation}
Set
\begin{equation}
\alpha_h(x)=x^{2h-1}{(h-1)!\over (h-1)!}F_h(x)\,.
\end{equation}
Then we have
\begin{equation}
{\alpha_h}^{(h)}(x)=x^{h-1}e^{-x}\,,
\end{equation}
(the superscript clearly denoting repeated differentiation).
Clearly we obtain
(up to terms involving constants of integration)
\begin{equation}
{\alpha_h}^{(h-1)}(x)=-\sum_{r=0}^{h-1}{(h-1)!\over (h-1-r)!}x^{h-1-r}
e^{-x}\,.
\end{equation}
Proceeding in this way, we find
\begin{equation}
{\alpha_h}^{(h-p)}(x)=(-1)^p\sum_{r_1=0}^{h-1}\sum_{r_2=0}^{h-1-r_1}
\cdots\sum_{r_p=0}^{h-1-r_1-\cdots-r_{p-1}}{(h-1)!\over
(h-1-r_1-\cdots-r_p)!}x^{h-1-r_1-\cdots-r_p} e^{-x}\,,
\end{equation}
or
\begin{equation}
\label{math225}
{\alpha_h}^{(h-p)}(x)=(-1)^p\sum_{r=0}^{h-1}(-1)^r\left({-p\atop
r}\right){(h-1)!\over (h-1-r)!}x^{h-1-r}e^{-x}\,.
\end{equation}
We thus find that, for some constants $q_{h,r}$,
\begin{equation}
\label{math12}
\alpha_h(x)=(-1)^h\sum_{r=0}^{h-1}(-1)^r\left({-h\atop r}\right)
{(h-1)!\over (h-1-r)!}x^{h-1-r}\left(e^{-x}-q_{h,r}\right)\,.
\end{equation}
We claim that setting $q_{h,r}=(-1)^{h-1-r}$
forces the coefficients
of $x^r$ for $0\leq r\leq h-1$ to vanish (which we certainly require).
This is easily
seen by differentiating the contribution from the integraion constants
in \reg{math12} and comparing with
\reg{math225}, and hence
$e^{{1\over 2}x}\alpha_h(x)$  is an odd function. Therefore,
$e^{{1\over 2}x}F_h(x)$ is even if and only if $h$ is even.

Thus, the general solution for $|P\rangle$ when $\chi$ is not constrained
to be a ground state is
\begin{equation}
\label{solution}
|P\rangle=|0\rangle+\sum_{\phi_i,n}\beta_{i,n}{L_{-1}}^n\phi_i\,,
\end{equation}
where $\beta_{i,n}$ are as given above (with arbitrary real $\beta_i$)
and the sum is over all real quasi-primary fields
$\phi_i$ of even weight $h_i$.

We are left with imposing orthogonality to $O(\Hil)$. However, note that
this has already been achieved for much of $O(\Hil)$, since
\begin{equation}
O(a,|0\rangle)=(L_{-1}+L_0)a\,,
\end{equation}
which is automatically orthogonal to $P$ by \reg{L1} (${L_n}^\dagger=
L_{-n}$).
We shall impose orthogonality in the form
\begin{equation}
\label{orthog}
\sum_{r=0}^h\left({h\atop r}\right)V(a)_{r+1}P=0\,,
\end{equation}
where $a$ is of conformal weight $h$. \reg{orthog} must be satisfied
for all $a\in\Hil$.
The above comment simply becomes
the observation that the term proportional to the vacuum $|0\rangle$
in \reg{orthog} vanishes.

We end this section with a few simple observations.
The trivial solution $P=|0\rangle$ to the above equations corresponds to
the adjoint representation.
When the state $\chi$ is
chosen to be an eigenstate of $L_0$, its conformal weight is given by
\begin{equation}
\label{weight}
\langle P|\psi_L\rangle\,,
\end{equation}
>from \reg{zeromode}. We also note that
given solutions $P_1$ and $P_2$ to the above equations,
then $\alpha P_1+(1-\alpha)P_2$ for all real $\alpha$ is also
a solution. It corresponds to the direct sum of the corresponding
representations.
Also, the tensor product of solutions $P_1$ and $P_2$
corresponding to (not necessarily identical) conformal field theories $\Hil_1$
and $\Hil_2$
corresponds to representations of $\Hil_1\oplus\Hil_2$.
\section{An Application to the Heisenberg Algebra}
\label{simple}
Let us introduce some notation. We follow \cite{DGMtwisted} and define
the FKS (untwisted) lattice conformal field theory as follows.
Let $\Lambda$ be an even Euclidean lattice of dimension $d$. Introduce
bosonic creation and annihilation operators $a_n^i$, $1\leq i\leq d$,
$n\in\ze$, such that
\begin{equation}
[a_m^i,a_n^j]=m\delta^{ij}\delta_{m,-n}\,,\qquad {a_m^i}^\dagger=a^i_{-m}\,.
\end{equation}
Set $a_0^i\equiv p^i$ and define $q^i$ by $[q^i,p^j]=i\delta^{ij}$.
The Hilbert space is built up by the action of the $a^i_{-n}$, $n>0$, on
momentum states $|\lambda\rangle$, $\lambda\in\Lambda$, such that
\begin{equation}
p^i|\lambda\rangle=\lambda^i
|\lambda\rangle\,,\qquad a^i_n|\lambda\rangle=0\,,
\end{equation}
for $n>0$.

The vertex operators are given by
\begin{equation}
\label{firstdef}
V(\phi,z)=:\prod_{a=1}^M{i\over (n_a-1)!}{d^{n_a}\over dz^{n_a}}
X^{i_a}(x) e^{i\lambda\cdot X(z)}:\sigma_\lambda\,,
\end{equation}
where
\begin{eqnarray}
\label{state}
\phi&=&\prod_{a=1}^Ma_{-n_a}^{i_a}|\lambda\rangle\,,\\
\label{stringfield}
X^i(z)&=&q^i-ip^i\ln z+i\sum_{n\neq 0}{a_n\over n}z^{-n}
\end{eqnarray}
and
\begin{equation}
\hat\sigma_\lambda\equiv e^{i\lambda\cdot q}\sigma_\lambda=
\sum_{\mu\in\Lambda}\epsilon(\lambda,\mu)|\mu+\lambda\rangle\langle\mu|\,,
\end{equation}
with
\begin{equation}
\label{sigma}
\hat\sigma_\lambda\hat\sigma_\mu=(-1)^{\lambda\cdot\mu}
\hat\sigma_\mu\hat\sigma_\lambda
\end{equation}
and $\epsilon(\lambda,\mu)=\pm 1$
suitably chosen \cite{DGMtwisted}.

This defines a conformal field theory $\Hil(\Lambda)$
of central charge $d$
($\psi_L={1\over 2}a_{-1}\cdot a_{-1}|0\rangle$), with conjugation
map given by
\begin{equation}
\overline\phi=(-1)^{L_0}\theta\phi\,,
\end{equation}
for $\phi$ as in \reg{state}, with
\begin{equation}
\theta a_n^i\theta^{-1}=-a^i_n\,,\qquad\theta|\lambda\rangle=|-
\lambda\rangle\,.
\end{equation}
The meromorphic representations of this are known
\cite{DGMtriality,dongtwisrep}.
They are simply given by the above vertex operators acting on the Hilbert
space of states generated by the bosonic creation operators acting on momentum
states $|\mu\rangle$, $\mu\in\lambda_0+\Lambda$, for $\lambda_0$ some fixed
element of $\Lambda^\ast$. [We must extend the definition of the $\epsilon(
\lambda,\mu)$ to $\mu\in\lambda_0+\Lambda$.]
That is, we obtain $|\Lambda^\ast/\Lambda|$
representations (and hence a unique meromorphic representation in the case
that $\Lambda$ is self-dual, {\em i.e.} the conformal field theory is
``self-dual''
-- see \cite{PSMreps,PGmer} for a further discussion of this).

Now consider just the Heisenberg algebra $H$ ({\em i.e.} $d=1$ and we set
all momenta to zero). The meromorphic representations then have Hilbert
spaces built up by the action of a single set of bosonic creation operators
on one-dimensional momentum states $|\mu\rangle$ for $\mu$ an arbitrary
real number. Since we require $\chi$ to be real we shall take
$\chi=i^{{1\over 2}\mu^2}|\mu\rangle_+/\sqrt 2\equiv
i^{{1\over 2}\mu^2}(|\mu\rangle+|-\mu\rangle)/\sqrt 2$ or $\chi=ii^{{1\over
2}\mu^2}
|\mu\rangle_-/\sqrt 2
\equiv i^{{1\over 2}\mu^2} i(|\mu\rangle-|-\mu\rangle)/\sqrt 2$.
The corresponding $P$ is then, in both
cases, found to be \cite{PSMreps,DGMtwisted}
\begin{equation}
\label{known}
\overline W(\chi,1)e^{L_1}|\chi\rangle=\rm{cosh}\,\left(\mu X_-\right)|0\rangle
\equiv P_\mu\,,
\end{equation}
where $X_-=\sum_{n=1}^\infty(-1)^n{a_{-n}\over n}$. (We ignore terms in the
sectors with momentum $\pm 2\mu$, since they will give no contribution
to the matrix element in \reg{maindef}. Note that this is an example
illustrating the comment on the definition of $P$ following
equation \reg{maindef}.) Let us try to derive this
result from our approach.

It turns out that in this simple case we do not need the general solution
for $P$ derived in the last section. We simply impose orthogonality to $O(H)$.
It is easily shown (Lemma 2.1.2 of \cite{Zhu}) that
$(a_{-n}+a_{-n-1})|0\rangle\in O(H)$, and we deduce that the most general
solution for $P$ orthogonal to $O(H)$ and satisfying the first two relations
of \reg{constraints} is
\begin{equation}
P=|0\rangle+\sum_{n=1}^\infty\lambda_n{X_-}^n|0\rangle\,,
\label{2star}
\end{equation}
where the $\lambda_n$ are arbitrary real coefficients. Trivially from
the general solution \reg{solution} for $P$, $\lambda_n=0$ for $n$ odd.
Since $L_{\pm 1}$ do not mix states with different numbers of creation
operators (in the zero-momentum sector) we see that we can get no further
information from this and that the solution we have is complete (equivalently
we could observe that since $P_\mu$ satisfies the constraint equations
for all $\mu$ then we can see that \reg{2star} (with  $\lambda_n=0$ for $n$
odd) must do so also).

To obtain the required form for $P$, we could impose that $\chi$ be
an
$L_0$ eigenstate by setting
\begin{equation}
\langle\chi|{L_0}^n|\chi\rangle=\langle\chi|L_0|\chi\rangle^n\,.
\end{equation}
But
\begin{eqnarray}
\langle\chi|{L_0}^n|\chi\rangle&=&\langle\chi|U(\psi_L)_0\ldots
U(\psi_L)_0|\chi\rangle\nonumber\\
&=&\langle\chi|U(\psi_L*\cdots *\psi_L)_0|\chi\rangle\nonumber\\
&=&\langle P|\psi_L*\cdots *\psi_L\rangle\,,
\end{eqnarray}
>from \reg{zeromode} and \reg{product}, and so we get
\begin{equation}
\label{eigenrel}
\langle P|\psi_L\rangle^n=\langle P|\psi_L*\cdots *\psi_L\rangle\,.
\end{equation}
This clearly fixes $\lambda_{2n}$ in terms of the lower order coefficients,
and so we see from the known solution that we must obtain $\lambda_{2n}
={\lambda^{2n}\over (2n)!}$ for some real $\lambda$.

In this case, we find that all solutions to our constraint equations
correspond to representations (reinforcing our belief that the conditions
are actually sufficient). Also note that, though there was no a priori
imposition that the representation be meromorphic (and non-meromorphic
representations certainly exist, as we shall see in the next section),
orthogonality to $O(H)$
seems to have restricted to meromorphic representations (some of the
manipulations we used
are strictly only valid in this case, so this is not too surprising).
It is also worth noting that the representation we obtain is not irreducible,
since it contains both that built up from $|\lambda\rangle$ and that
>from $|-\lambda\rangle$, though the distinct solutions we have found
do correspond to inequivalent representations as can be trivially seen by the
fact that the conformal weight of the ground state is distinct.
In general though, we will have to do more work
after solving our equations to identify the inequivalent as well as the
irreducible representations of a
given theory.
\section{The $\ze_2$-Twisted Heisenberg Algebra and the Reflection-Twisted
FKS Lattice Conformal
Field Theory}
\label{notsosimple}
\subsection{Notation and known results}
We begin by introducing a non-meromorphic representation of the conformal
field theory $\Hil(\Lambda)$ (see \cite{DGMtwisted} for details).
We start with an irreducible representation
of the gamma matrix algebra
\begin{equation}
\gamma_\lambda\gamma_\mu=(-1)^{\lambda\cdot\mu}\gamma_\mu
\gamma_\lambda\,,
\end{equation}
{\em c.f.} \reg{sigma}. We denote a typical state in such a representation
by $\rho$. The Hilbert space for our representation of $\Hil(\Lambda)$
is built up from this space by the action of bosonic creation oscillators
with non-integral grading, {\em i.e.} we introduce operators
$c^i_r$, $1\leq i\leq d$, $r\in\ze+{1\over 2}$, such that
\begin{equation}
[c^i_r,c^j_s]=r\delta^{ij}\delta_{r,-s}\,,\qquad {c^i_r}^\dagger
=c^i_{-r}\,,
\end{equation}
and
\begin{equation}
c^i_r\rho=0
\end{equation}
for $r>0$. The ``twisted'' vertex operator (the $U$ of our general theory)
for $\phi$ as in
\reg{state} is given by \cite{DGMtwisted}
\begin{equation}
\label{mystery}
V_T(\phi,z)=V_0\left(e^{\Delta(z)}\phi,z\right)\,,
\end{equation}
where
\begin{equation}
\label{twistedvops}
V_0(\phi,z)=:\prod_{a=1}^M{i\over (n_a-1)!}{d^{n_a}\over dz^{n_a}}
R^{i_a}(x) e^{i\lambda\cdot R(z)}:\gamma_\lambda\,,
\end{equation}
with
\begin{equation}
R^i(z)=i\sum_{r\in\ze+{1\over 2}}{c_r\over r}z^{-r}
\end{equation}
and
\begin{equation}
\label{delooo}
\Delta(z)=-{1\over 2}p^2 \ln 4z + {1\over 2}\sum_{{m,n\geq 0\atop
(m,n)\neq (0,0)}}
\left({-{1\over 2}\atop m}\right)\left({-{1\over 2}\atop n}\right)
{z^{-m-n}\over m+n} a_m\cdot a_n\,.
\end{equation}
This is found to define a representation of $\Hil(\Lambda)$, which we
denote $\Hil_T(\Lambda)$, the so-called $\ze_2$-twisted
representation. Note that it is non-meromorphic however. The matrix
elements of the $V_T$'s contain square
root branch cuts in general.
Also note that the conformal weight of the ground state of the
representation is
found to be $d/16$.
The involution $\theta$ defined on $\Hil(\Lambda)$ can be lifted to this
representation by
\begin{equation}
\theta\rho=\rho\,,\qquad\theta c^i_r\theta^{-1}=-c^i_r\,.
\end{equation}
[Note that this is gives an involution on the representation
when $d$ is a multiple of 8 if we define instead $\theta\rho=(-1)^{d/8}
\rho$.]

We set
\begin{equation}
\Hil_T(\Lambda)_\pm=\left\{\zeta\in\Hil_T(\Lambda)\ :\ \theta\zeta=
\pm\zeta\right\}\,,
\end{equation}
with a similar decomposition for $\Hil(\Lambda)$.
Then $\Hil(\Lambda)_\pm$ and $\Hil_T(\Lambda)_\pm$ are found to form
meromorphic irreducible
representations of $\Hil(\Lambda)_+$ \cite{DGMtriality}.
Our main result will be that these are the only such representations,
at least for $\Lambda$ self-dual.

Let us now, as before, restrict to one dimension and set all momenta
to zero, {\em i.e.} we will study representations of $H_+$,
the $\theta=1$
projection of the Heisenberg algebra $H$.

Let us look at the known results before we start to analyze solutions
of our equations. From the above, we have meromorphic
representations corresponding to real ground states ({\em i.e.} possible
choices for $\chi$),
\begin{eqnarray}
\label{setstates}
\chi_1=i^{{1\over 2}\lambda^2}|\lambda\rangle_+\,,\quad
\chi_2&=&ii^{{1\over 2}\lambda^2}|\lambda\rangle_-\,,\quad
\chi_3=a_{-1}|0\rangle\,,
\nonumber\\
\chi_4=\rho\,,&\quad&
\chi_5=c^i_{-{1\over 2}}\rho'\,,
\end{eqnarray}
for suitable $\rho$ and $\rho'$ (see \cite{DGMtwisted} for a discussion of the
action
of conjugation
on the twisted sector ground states).
(The term $\chi_3$ arises since $|0\rangle_-=0$.) We now calculate the
corresponding $P$'s. As before,
$\chi_1$ and $\chi_2$ give rise to $P_\lambda$ \reg{known}, again
ignoring contributions to $P$ with non-zero momentum. The other
three $\chi$'s give us three exceptional solutions $P_a$, $P_b$ and $P_c$
corresponding to ground states of conformal weight 1, ${1\over 16}$ and
${9\over 16}$ respectively.
We evaluate them to be
\begin{eqnarray}
\label{known2}
P_a&=&|0\rangle-\sum_{n=1}^\infty (-1)^na_{-1}a_{-n}|0\rangle\nonumber\\
P_b&=&e^{\Delta(1)^\dagger}|0\rangle\nonumber\\
P_c&=&\left(1-2\sum_{n,m>0}\left({{1\over 2}\atop m}\right)\left({-{1\over
2}\atop n}
\right)a_{-m}^ia_{-n}^i\right)e^{\Delta(1)^\dagger}|0\rangle\,.
\end{eqnarray}
\subsection{Solution in a simple case}
Let us start with a simple exercise to see how these solutions corresponding to
particular
values of the conformal weight of the representation ground state might arise.

We shall look for solutions for $P$ corresponding to representations of $H_+$
in which
$P$ has no more than two creation operators in any one term. From the above
results
and our conjecture of their completeness, we would expect only to obtain the
trivial
solution $|0\rangle$ and $P_a$.

We first write down a general solution to the constraint equations
\reg{constraints}
using \reg{solution}, and then impose orthogonality to $O(H_+)$.
The quasi-primary states in $H_+$ containing two oscillators are easily found.
They are,
at even levels $2h$,
\begin{equation}
\phi_h=\sum_{r=1}^h{(-1)^r\over r}\left({2h-1\atop r-1}\right)P_{r,2h-r}
-{1\over 2}{(-1)^h\over h}\left({2h-1\atop h-1}\right)P_{h,h}\,,
\end{equation}
where $P_{n_1n_2\ldots}\equiv a_{-n_1}a_{-n_2}\cdots|0\rangle$, and so we write
\begin{equation}
P=|0\rangle+\sum_{h=1}^\infty\sum_{n=0}^\infty\beta(2h,n)\alpha_h{L_{-1}}^n\phi_h\,,
\end{equation}
for some real coefficients $\alpha_h$.
We now impose \reg{eigenrel}, {\em i.e.} we take $\chi$ to be a conformal
eigenstate.
This clearly gives us $\alpha_h$ in terms of $\alpha_1$. Set
$\alpha_1=-2\gamma$
($\gamma$ will be the conformal weight of the representation (from
\reg{weight}).
Explicitly we find
\begin{eqnarray}
\alpha_2&=&-{1\over 3}\gamma\left(\gamma+{1\over 5}\right)\nonumber\\
\label{alpha3}
\alpha_3&=&-{1\over 45}\gamma\left(\gamma^2+\gamma+{1\over 7}
\right)\,,
\end{eqnarray}
and, up to and including states of conformal weight 6,
\begin{eqnarray}
P&=&|0\rangle+\gamma P_{11}-\gamma P_{12}+{1\over 3}\gamma(\gamma+2)
P_{13}-{1\over 4}\gamma(\gamma-1)P_{22}-{1\over 3}\gamma(\gamma-1)P_{23}
-\nonumber\\&&{1\over 2}\gamma(\gamma+1)P_{14}+{\gamma\over 45}\left(
\gamma^2+26\gamma+18\right)P_{15}-{\gamma\over 36}\left(2\gamma^2+7\gamma
-9\right)P_{24}+\nonumber\\
&&{\gamma\over 27}\left(\gamma^2-4\gamma+3\right)P_{33}\,.
\end{eqnarray}
Now, ignoring terms with more than two oscillators,
\begin{equation}
O(P_{11},P_{12})=2P_{22}+4P_{13}+4P_{23}+8P_{14}+2P_{24}+4P_{15}\,,
\end{equation}
and so setting this orthogonal to $P$ we find
\begin{equation}
\gamma^3+35\gamma^2-36\gamma=0\,.
\end{equation}
This has roots $\gamma=0$, $1$ as required, as well as the root
$\gamma=-36$ which we can dismiss by unitarity (though it is spurious and we
will see from
the argument that we use in the general case that $\gamma=0$, 1 are the only
possible
solutions). $\gamma=0$ gives $P=|0\rangle$, while $\gamma=1$ must give
$P=P_a$, since this solution must arise from this analysis (we
evaluate the first few terms
as a check on our techniques).
\subsection{General solution for $H_+$}
Let us now attempt the general case. Instead of imposing \reg{eigenrel}, we
shall require the (more restrictive) condition, using \reg{product},
$\langle P|\psi_L*\phi\rangle=\gamma\langle P|\phi\rangle$ for all $\phi\in
H_+$ and
with $\gamma$ the required conformal weight of the ground state (the
coefficient of
$P_{11}$ in $P$ in this case). Since $\phi$ is arbitrary, this condition
amounts to
requiring
\begin{equation}
(L_2-L_0)P=\gamma P\,,
\label{L2reln}
\end{equation}
using \reg{L1} and the definition of $*$ in \reg{stardef}.
This is easier to use in practise than \reg{eigenrel}, and it turns out is also
more restrictive (since we are using \reg{product}, which itself depends on
$\chi$ being
a ground state in the representation). Note that \reg{L2reln} together with
\reg{solution} will give us a solution for $P$ up to arbitrary coefficients for
the
primary states at even levels. Since there are no primary states involving just
two
oscillators, then $P$ in the above simple example would be determined
completely in terms
of $\gamma$ (so we see immediately that \reg{L2reln} is strictly more powerful
than \reg{eigenrel}). However, in the general case, an infinite number of
primary states
occur and we obtain a corresponding set of
unknown coefficients which must be constrained by imposing \reg{orthog}.

We know from the Kac determinant \cite{Kac} that the even levels at which
primary
states occur are given by $4n^2$, $n=0,1,\ldots$. So, we expect to obtain at
least
one new parameter at level 4 (in fact exactly one, but we shall postpone a
discussion of the explicit structure of the primary states until it is needed
later in the argument). We start from \reg{solution}, and so must consider
the quasi-primary states. Since $L_1$ does not mix terms with different
numbers of oscillators, we can consider quasi-primary states at a given
level and with a certain number of oscillators. The two oscillator
ones $\phi_h$ at level $2h$ are as given above. We denote their coefficients
in the expansion \reg{solution} of $P$ as $\gamma_h$, and set as before
$\gamma_1=-2\gamma$, $\gamma$ the conformal weight of the ground state of the
representation. We find a 4-oscillator quasi-primary
state at level 4, which is simply $P_{1111}$.  We denote the coefficient of
this
in the expansion \reg{solution} of $P$ as $\rho$. At level 6 we have, in
addition to
the quasi-primary state with two oscillators, one 4-oscillator one
\begin{equation}
3P_{1122}-4P_{1113}\,,
\end{equation}
whose coefficient we denote by $\delta$,
and one trivial 6-oscillator state $P_{111111}$.

Imposing \reg{L2reln}, we find
\begin{eqnarray}
\gamma_2&=&2\rho-{1\over 3}\gamma\left(\gamma+{1\over 5}\right)\nonumber\\
45\gamma_3&=&{9\over 5}\rho\left({8\over 3}+4\gamma\right)-{6\over 5}\gamma^3-
{4\over 5}\gamma^2-{1\over 7}\gamma\nonumber\\
\delta&=&-{\gamma^2\over 180}+{\gamma^3\over 180}-{\rho\over 45}
-{\gamma\rho\over 30}\,.
\end{eqnarray}
Note that the expression for $\gamma_3$ differs from that in
\reg{alpha3} when we take $\rho=0$, but we are now using the more powerful
relation \reg{L2reln} and the expressions in any case then agree for the
values $\gamma=0$, 1 which is all we can really require.
Note also that if we set $\delta=\rho=0$ ({\em i.e.} again restrict to at most
two oscillators) then we again find $\gamma=0$, 1 (this time without the
spurious negative root).

Now that we have some experience with this technique, let us consider what
exactly
the result is which we are trying to find. If we are to find $P_\lambda$,
$P_a$, $P_b$
and $P_c$ as the only possible solutions, then we trivially see we must have
$\rho={\gamma^2\over 6}$ except at $\gamma=1$, ${1\over 16}$ and ${9\over 16}$.
(We expect there to be no constraint on $\rho$ when $\gamma$ is one of the
special values listed, since in that case we expect $\alpha
P_\lambda+(1-\alpha)P_x$
($x=a$, $b$ or $c$) to be a solution for all real $\alpha$.)
The lowest degree polynomial which will provide such a relation is of degree
5 in $\gamma$. We see, for the techniques we are using, that this can come only
>from a state of conformal weight at least 10.
Since the computation grows rapidly with increasing conformal weight, we hope
that
the required level is exactly 10.
It would be expected that some new feature occurs at this level.

Let us look at the numbers of quasi-primary states at the various levels. From
the
partition function
\begin{equation}
1+x^2+x^3+3x^4+3x^5+6x^6+7x^7+12x^8+14x^9+22x^{10}+\ldots\,,
\end{equation}
we see that there are 5 quasi-primary states at level 8 and 8 at level 10.
We can construct a set of quasi-primary states from the known two-oscillator
ones.
For example, at level 4 we can write the two quasi-primaries as $\phi_2$ and
${\phi_1}^2$ (using an obvious notation -- more precisely we are projecting
the $*$ product $\phi_1*\phi_1$ onto the state of highest conformal weight).
We find that this gives all quasi-primary states at levels 2, 4, 6 and 8, but
at level 10 we have to consider in addition the quasi-primary state
$P_{1144}-4P_{1234}+{64\over 27}P_{1333}+2P_{4222}-{4\over 3}P_{2233}$,
which is the required new feature.
Imposing \reg{L2reln} on \reg{solution} then fixes all unknown parameters in
$P$ up
to level 10 in terms of $\rho$ and $\gamma$ (since the next primary state is at
level
16).
We find that this imposition is consistent, and we must consider \reg{orthog}
if
we are to get our required constraint.

The first non-trivial state to try in \reg{orthog} should be $P_{22}$, since
the action of the Virasoro algebra on $P$ is determined by the action of $L_2$
and $L_1$ (which have been considered exhaustively),
and the first state not in the Virasoro
module on the vacuum is $P_{22}$.
We find that
\begin{equation}
[V(P_{22})_m,a_p]=-2p(1+m+p)(1-p)a_{p+m}\,.
\end{equation}
Acting upon our expansion of $P$, we find that all terms vanish up to and
including level 4, but at level 5 we obtain some potentially
non-zero coefficients.
Equating the coefficient of $P_{1112}$ to zero gives
\begin{equation}
(9-169\gamma+416\gamma^2-256\gamma^3)\left(\rho-{\gamma^2\over 6}\right)
=0\,,
\end{equation}
which is the required result.

All that remains now is to argue that the terms at higher levels in $P$ are
given uniquely in terms of those which we have already computed. This unique
form must then be the same as the known solutions.

We consider the primary states of the theory. We have a specific form for these
>from \cite{Jim'sBRST,KacRaina,WakimotoPrimary}. We find that the primary
fields
occur at level $n^2$, $n\in\ze$, and
that the unique \cite{KacRaina,WakimotoPrimary} primary at level
$h=n^2$ is given by
\begin{equation}
\label{primary}
S_{\underbrace{n,\ldots,n}_{{\rm n}\ {\rm terms}
}}\left(\sqrt 2 a_{-j}/j\right)|0\rangle\,,
\end{equation}
where
the Schur polynomial associated to a partition $\lambda=\{
\lambda_1\geq\lambda_2\geq\ldots\}$ is
\begin{equation}
S_{\lambda_1,\lambda_2,\ldots}(x)={\rm det}\,\left(S_{\lambda_i+j-i}(x)
\right)_{i,j}\,,
\end{equation}
and the elementary Schur polynomials $S_k(x)$ are defined by
\begin{equation}
\sum_{k\geq 0}S_k(x)z^k=\exp\left(\sum_{k\geq 1}x_kz^k\right)\,.
\end{equation}
Let us show, for reasons that will become clear, that the coefficient
of ${a_{-1}}^{n^2}|0\rangle$ in the primary field is non-zero.
It is easily seen to be
\begin{equation}
\det\left(1/(n+j-i)!\right)\,,
\end{equation}
and this elementary determinant we find, for example
>from \cite{FrobeniusExample}, to be
\begin{equation}
\label{endprimary}
{\prod_{i=1}^n i!\over\prod_{i=n}^{2n-1}i!}\,,
\end{equation}
giving the required result.

Rather than use the result for the primary field given in
\reg{primary}, we find it easier to construct the terms we require for
our argument explicitly. We begin by writing down the quasi-primary states
at level $h$ with$h$, $h-2$, $h-4$
and $h-6$ oscillators.
They are found to be
\begin{eqnarray}
\psi_1&=&{a_{-1}}^h|0\rangle\nonumber\\
\psi_2&=&4{a_{-1}}^{h-3}a_{-3}|0\rangle-3{a_{-1}}^{h-4}{a_{-2}}^2
|0\rangle\nonumber\\
\psi_3&=&3{a_{-1}}^{h-8}{a_{-2}}^4|0\rangle+{16\over 3}{a_{-1}}^{h-6}
{a_{-3}}^2|0\rangle-8{a_{-1}}^{h-7}{a_{-2}}^2a_{-3}|0\rangle\nonumber\\
\psi_4&=&2{a_{-1}}^{h-5}a_{-5}|0\rangle-5{a_{-1}}^{h-6}a_{-2}a_{-4}
|0\rangle+{10\over 3}{a_{-1}}^{h-6}{a_{-3}}^2|0\rangle\nonumber\\
\psi_5&=&{a_{-1}}^{h-12}{a_{-2}}^6|0\rangle-{64\over 27}{a_{-1}}^{h-9}
{a_{-3}}^3|0\rangle+{16\over 3}{a_{-1}}^{h-10}{a_{-2}}^2{a_{-3}}^2|0\rangle
-4{a_{-1}}^{h-11}{a_{-2}}^4a_{-3}|0\rangle\nonumber\\
\psi_6&=&{a_{-1}}^{h-7}a_{-7}|0\rangle-{7\over 2}{a_{-1}}^{h-8}a_{-2}
a_{-6}|0\rangle+7{a_{-1}}^{h-8}a_{-3}a_{-5}|0\rangle-{35\over 8}
{a_{-1}}^{h-8}{a_{-4}}^2|0\rangle\nonumber\\
\psi_7&=&{a_{-1}}^{h-9}{a_{-2}}^2a_{-5}|0\rangle-{5\over 3}{a_{-1}}^{h-9}
a_{-2}a_{-3}a_{-4}|0\rangle+{20\over 27}{a_{-1}}^{h-9}{a_{-3}}^3|0\rangle
-{4\over 3}{a_{-1}}^{h-8}a_{-3}a_{-5}|0\rangle\nonumber\\
&&\hskip20pt +{5\over 4}{a_{-1}}^{h-8}
{a_{-4}}^2|0\rangle\nonumber\\
\psi_8&=&{1\over 2}{a_{-1}}^{h-8}{a_{-4}}^2|0\rangle-2{a_{-1}}^{h-9}
a_{-2}a_{-3}a_{-4}|0\rangle+{32\over 27}{a_{-1}}^{h-9}{a_{-3}}^3
|0\rangle-{2\over 3}{a_{-1}}^{h-10}{a_{-2}}^2{a_{-3}}^2|0\rangle\nonumber\\
&&\hskip20pt +{a_{-1}}^{h-10}{a_{-2}}^3a_{-4}|0\rangle\,.
\end{eqnarray}
We then impose that the primary state is annihilated by $L_2$.
This determines the coefficients $\epsilon_n$ of $\psi_n$ in terms
of $\epsilon_1$, which we take (in the light of the above
analysis) to be 1 for convenience.
We find in particular
\begin{eqnarray}
\epsilon_3&=&{1\over 384}h(h-1)(h-4)(h-21)\nonumber\\
\epsilon_4&=&{1\over 20}h(h-1)(h-4)\nonumber\\
\epsilon_6&=&-{5\over 36}h(h-1)(h-4)(h-9)\,.
\end{eqnarray}
We consider the action of $V\left(P_{22}\right)_5$ on
the primary state at level $h=n^2$. If we can show that this is non-zero, then
the relation \reg{orthog} for $a=P_{22}$ will give the coefficient
of the primary state in terms of the coefficients of states at lower
conformal weight.
We will consider the term
${a_{-1}}^{h-7}a_{-2}|0\rangle$ which arises as a result of the action.
The relevant pieces of $V\left({a_{-2}}^2|0\rangle\right)_5$ are
\begin{equation}
24a_2a_3+20a_1a_4-16a_{-2}a_7\,,
\end{equation}
and we need to consider their action on the states
${a_{-1}}^{h-7}a_{-7}|0\rangle$, ${a_{-1}}^{h-7}{a_{-2}}^2a_{-3}
|0\rangle$ and ${a_{-1}}^{h-6}a_{-2}a_{-4}|0\rangle$. Assign these
coefficients $y_1$, $y_2$ and $y_3$ respectively.
We find trivially that the coefficient of ${a_{-1}}^{h-7}a_{-2}|0\rangle$
will be
\begin{equation}
-112y_1+288y_2+80(h-6)y_3\,,
\end{equation}
and substituting in our results for the primary state we get
\begin{equation}
-4h(h-1)(h-4)(4h-39)\,,
\end{equation}
which does not vanish over the range of relevant values for $h$.

This completes the proof that the only possible solutions for $P$ for the
algebra $H_+$ are $P_\lambda$, $P_a$, $P_b$ and $P_c$. That each correspond to
consistent meromorphic representations is known from the explicit results of
previous work.
\subsection{General solution for $\Hil_+(\Lambda)$, $\Lambda$ one-dimensional}
We now consider representations of the theory $\Hil_+(\Lambda)$ in the
case where $\Lambda$ is a one-dimensional lattice.

We already have the following $P's$ ({\em c.f.} \reg{known} and
\reg{known2})
\begin{eqnarray}
\hat P^\pm_\mu&\equiv&\cosh (\mu X_-) \left(|0\rangle\pm\sqrt 2|2\mu\rangle_+
\right)\nonumber\\
\hat P_a&\equiv&P_a\nonumber\\
\hat P_b&\equiv&\sum_{\lambda\in\Lambda}\rho^\dagger\gamma_\lambda\rho
e^{\Delta(1)^\dagger}|\lambda\rangle
\equiv{1\over 2}\sum_{\lambda\in\Lambda_0}S_\lambda e^{
\Delta(1)^\dagger}|\lambda\rangle_+\nonumber\\
\hat P_c&\equiv&\sum_{\lambda\in\Lambda}\rho^\dagger\gamma_\lambda\rho
\left(1-2\sum_{n,m\geq 0}\left({{1\over 2}\atop m}\right)\left({-{1\over
2}\atop n}
\right)a_{-m}^ia_{-n}^i\right)e^{\Delta(1)^\dagger}|\lambda\rangle\nonumber\\
&\equiv&{1\over 2}\sum_{\lambda\in\Lambda_0}S_\lambda
\left(1-2\sum_{n,m\geq 0}\left({{1\over 2}\atop m}\right)\left({-{1\over
2}\atop n}
\right)a_{-m}^ia_{-n}^i\right)e^{\Delta(1)^\dagger}|\lambda\rangle_+
\,,
\end{eqnarray}
where $\Lambda_0=\{\lambda\in\Lambda:{1\over 2}\lambda^2\ {\rm even}\}$,
corresponding to representation states as before ({\em i.e.} \reg{setstates}),
which are easily evaluated from the known representations detailed earlier.
(Note though that $|\mu\rangle_\pm$ is only a ground state for a meromorphic
representation if $\mu$ is a vector of minimal norm in the cosets
$\Lambda^\ast/\Lambda$, and of course the term $|2\mu\rangle_+$ is present
only if $2\mu\in\Lambda$.)
[In one dimension, the cocycle structure is trivial, and if we take
$\lambda_0$ to be a basis vector for $\Lambda$ then the only
(irreducible)
representations (with $\gamma_\lambda=\gamma_{-\lambda}=
(-1)^{{1\over 2}\lambda^2}{\gamma_\lambda}^\dagger$ and also for
which a suitable charge conjugation matrix exists \cite{DGMtwisted})
are one or two-dimensional and are such that
$S_{n\lambda_0}$ is either 1 or $(-1)^n$ in the case
that ${1\over 2}{\lambda_0}^2$ is even, and is $-{1\over 2}(1+(-1)^n)$
when ${1\over 2}{\lambda_0}^2$ is odd (so that $S_\lambda\equiv -1$
on $\Lambda_0$), giving us the above results.]

Let us now verify our uniqueness conjecture that these are the only solutions
corresponding to meromorphic representations.
We essentially follow the argument of the preceding subsection (though the
algebra is considerably more intricate), and so shall simply sketch the
proof.

We shall evaluate the terms in $P$ with momentum $\pm\lambda$, $\lambda\in
\Lambda$. As before, we begin by writing down the quasi-primary states
at even levels and expand $P$ as in \reg{solution}.
Imposing \reg{L2reln} then fixes, as before, all of the unknown coefficients
except for those of primary states at even levels.
 From the known results and our uniqueness conjecture, we would expect to
derive the constraint $\gamma={1\over 16}$, $\gamma={9\over 16}$ or
$\lambda^2=8\gamma$ (the last of these corresponding to $\hat
P_\mu$). Thus we expect a cubic equation in $\gamma$, and would thus
naively expect to have to evaluate $P$ as far as 6 levels above
$|\lambda\rangle_+$.

We have explicit forms for the primary states,
>from {\em e.g.} \cite{Jim'sBRST}, which tells us (remembering that
the momentum $\lambda$ is constrained so that $\lambda^2$ is even)
that the primary states are at levels $k(k+2 d)$ above
$|\sqrt 2 d\rangle_+$ for $k$ and $d$ non-negative integers.
In addition, the conformal weights of the primaries must be even
for an allowable contribution to $P$.

We demonstrate below that
the contribution from a primary state at level at least 7 above
$|\sqrt 2 d\rangle_+$ is determined in terms of the coefficients of
the primary states at lower levels (just as we did to complete the proof
in the case of zero momentum in the above subsection). This, together
with the known (even) levels of the primary states is easily seen to
give that for $d$ odd and at least 3 all contributions to $P$ vanish,
while for $d$ even everything is determined in terms of the coefficient
of
the primary state $|\sqrt 2 d\rangle_+$.

The case $d=1$ must be considered
separately. There is a primary state 3 levels above $|\sqrt 2\rangle_+$
(then the next is at level 15). This state is explicitly
\begin{equation}
{a_{-1}}^3|\sqrt 2\rangle_- - {3\over \sqrt 2} a_{-1}a_{-2}|\sqrt 2\rangle_+
+ a_{-3}|\sqrt 2\rangle_-\,,
\end{equation}
and a simple calculation confirms that it is not annihilated by
$V(|\sqrt 2>_+)_2$, and hence \reg{orthog} shows that its coefficient in $P$
must vanish. Then the argument below that higher primary states are determined
in terms of the lower ones shows that there are no contributions to $P$
corresponding to $d=1$.

We now give the argument that the higher order primary states are determined
as described above. As in (\ref{primary}-\ref{endprimary}), we find that the
coefficient of the term $a_{-1}^h|\lambda\rangle_\pm$ is non-zero in the
primary state at level $h$ above the state $|\lambda=\sqrt 2 d\rangle_+$
(with $h=k(k+2d)$). We then construct the primary state by hand in terms
of states of successively decreasing numbers of oscillators, as before.
We omit the details as, though the technique is straightforward, the explicit
forms involved are unwieldy.  As in the zero momentum case, we evaluate the
action of $V(P_{22})_5$ on the primary. In particular, we consider the term
${a_{-1}}^{h-5}|\lambda\rangle_\mp$ (for which we require the coefficients
$r$, $s$ and $t$
of the states ${a_{-1}}^{h-5}a_{-2}a_{-3}|\lambda\rangle_\mp$,
${a_{-1}}^{h-4}a_{-4}|\lambda_\mp$ and ${a_{-1}}^{h-5}
a_{-5}|\lambda\rangle_\pm$). We find that the coefficient of
${a_{-1}}^{h-5}|\lambda\rangle_\mp$ is
\begin{equation}
\label{nasty1}
144 r + 80 (h-4) s + 60 \lambda t
=4 \sqrt 2 h d (110 - 89 h + 12 h^2  + 157 d^2  - 40 h d^2  + 12 d^4 )\,.
\end{equation}
To eliminate the (presumably) spurious roots to the vanishing of this,
we consider also the coefficient of the term
${a_{-1}}^{h-7}|\lambda\rangle_\mp$ in the action of $V(P_{33})_5$
on the primary. This turns out to be
\begin{equation}
72(14 a + 35 b + 50 c + 7 \lambda d)\,,
\end{equation}
where $a$, $b$, $c$ and $d$ are the coefficients of
${a_{-1}}^{h-6}a_{-6}|\lambda\rangle_\mp$,
${a_{-1}}^{h-7}a_{-2}a_{-5}|\lambda\rangle_\mp$,
${a_{-1}}^{h-7}a_{-3}a_{-4}|\lambda\rangle_\mp$ and
${a_{-1}}^{h-7}a_{-7}|\lambda\rangle_\pm$
in the primary state respectively. This we evaluate to be
\begin{equation}
\label{nasty2}
12\sqrt 2 h d (5181 - 4737 h + 885 h^2  - 33 h^3  + 12167 d^2  -
 4945 h d^2  + 316 h^2  d^2  + 2824 d^4  - 364 h d^4  + 48 d^6 )\,,
\end{equation}
and eliminating $h$ between the vanishing of
\reg{nasty1} and \reg{nasty2} gives
\begin{eqnarray}
2304&&(d-1)(d+1)(2d-3)(2d+3)(2d-1)(2d+1)
   (-14402773 + 38418179 d^2 - \nonumber\\
&&18390543 d^4 +
     1290780 d^6)=0\,,
\end{eqnarray}
which has no integer solutions except $d=1$.
In the case $d=1$, the only integer solution for $h$ to \reg{nasty1} is
$h=3$, which is the state we considered separately above. This completes
our argument.

We may now evaluate the contributions to $P$ for states of momentum $\pm
\lambda$ (${1\over 2}\lambda^2$ even) as far as level 6 (up to
 which there are
no primary states other than $|\lambda\rangle_+$ itself).
There is one quasi-primary state at level 2 above $|\lambda\rangle_+$,
two at level 4 and 4 at level 6 to consider. The arguments are
exactly as before. We obtain a
consistent solution which we do not detail here since its exact structure
is not illuminating and the expressions involved are again unwieldy.
Following the preceding analysis with zero momentum, we then impose
\reg{orthog} with $a=P_{22}$. The coefficient of the term
$a_{-1}|\lambda\rangle_-$ is found to be
\begin{equation}
\label{gfhr}
32\lambda {(\gamma-{1\over 16})(\gamma-{9\over 16})(\lambda^2-8\gamma)
(\lambda^2-2)\over (2\lambda^2-1)^2(4\lambda^4-6\gamma\lambda^2+225)}
C_\lambda\,,
\end{equation}
where $C_\lambda$ is the coefficient of $|\lambda\rangle_+$ in $P$.
We thus find the required result.

All that remains to fix the form of $P$ exactly is to evaluate the
coefficients $C_\lambda$. The relation \reg{orthog} with $a=|\lambda_0
\rangle_+$ ($\lambda_0$ a basis vector for the lattice, as above)
will give a relation between $C_{\lambda+\lambda_0}$ and
$C_{\lambda-\lambda_0}$. The coefficient of $C_{\lambda+\lambda_0}$ in this
will be non-zero if the difference of the conformal weights of
$|\lambda\rangle_+$ and $|\lambda-\lambda_0\rangle_+$ is at least
${1\over 2}{\lambda_0}^2+1$ (the highest weight in \reg{orthog}). This
is true, for $\lambda=n\lambda_0$, if $n>1$. Thus, we see that we can determine
every coefficient $C_\lambda$ in terms of $C_0$ (which is fixed equal to 1
by \reg{constraints}) and $C_{\lambda_0}$. The arbitrariness of $C_{\lambda_0}$
(if ${1\over 2}{\lambda_0}^2$ is even, {\em i.e.} if $\lambda_0\in
\Lambda_0$ -- otherwise $C_{\lambda_0}=0$ as the primary state
$|\lambda_0\rangle_+$ is of odd conformal weight)
reflects the fact that we are able, in the case of the twisted representations,
to take a linear combination of known
solutions with $S_{n\lambda_0}=1$ and $S_{n\lambda_0}=(-1)^n$ ({\em i.e.}
the solution to the equations for $P$ does not correspond to
an irreducible representation, as we have
discussed before ). Similarly, in the case of a representation based on a
momentum state $|\mu\rangle_\pm$, we expect the coefficient of
$|2\mu\rangle_+$ (when $2\mu\in\Lambda$) to be arbitrary, because we
obtain an analogous linear combination of $\hat P^\pm_\mu$.

Note that, in order to constrain $\mu\in\Lambda^\ast$, we cannot use
\reg{orthog}, but must evaluate for example
the four-point function
\begin{equation}
\label{4pfn}
\langle\mu|V(|\lambda\rangle_+,z)V(|\lambda\rangle_+,w)|\mu\rangle
\end{equation}
This is easily done (see {\em e.g.} \cite{DGMtriality}), and we find
terms of the form $z^{\pm\lambda\mu}$. The restriction that the
representation be meromorphic gives the required constraint. Note
that this is our first example of a case in which solving all of our
equations for $P$
does not automatically give a meromorphic representation.
\subsection{Extension to $d$ dimensions}
In this section we shall describe the extension of the above results to the
case of $d$ dimensions. We only sketch the main arguments, since the
techniques used are straightforward though messy. In any case, such arguments
may clearly be made rigorous if desired.

Let us first consider the case of the $\ze_2$ projection ${H^d}_+$ of
the Heisenberg algebra in $d$ dimensions.

Now, in more than one dimension, it can be shown
\cite{DGMtriality} that the theory
${H^d}_+$ is generated by the modes of the vertex operators corresponding
to the states $a_{-1}^ia_{-1}^j|0\rangle$, $1\leq i\leq j\leq d$. Thus,
we see that we shall only need to consider the matrix elements involving
such states in order to fix the representation uniquely.
In practice, it turns out to be more convenient however to take the
following (though closely related) approach.

We first note that
\begin{equation}
\prod_{a=1}^Ma^{i_a}_{-m_a}|0\rangle*\prod_{b=1}^N
a^{j_b}_{-n_b}|0\rangle=\prod_{a=1}^Ma^{i_a}_{-m_a}
\prod_{b=1}^N
a^{j_b}_{-n_b}|0\rangle+\ldots\,,
\end{equation}
where $\ldots$ denotes terms containing less than M+N oscillators.
We see that by repeating this process we may thus write any state
in ${H^d}_+$ as a sum of $*$ products of states of the
form $a^i_{-m}a^j_{-n}|0\rangle$.
Hence we find that we need only consider
matrix elements of the form
\begin{equation}
\label{dooodah}
\langle\chi|U(a^{i_1}_{-m_1}a^{j_1}_{-n_1}|0\rangle*
\ldots a^{i_N}_{-m_N}a^{j_N}_{-n_N}|0\rangle)_0|\chi\rangle=
\langle\chi|U(a^{i_1}_{-m_1}a^{j_1}_{-n_1}|0\rangle)_0\ldots
U(a^{i_N}_{-m_N}a^{j_N}_{-n_N}|0\rangle)_0|\chi\rangle
\,,
\end{equation}
using \reg{product}.

Let us restrict now to the case $d=2$ for simplicity of notation.
Relabel the oscillators $a^1$ and $a^2$ as $a$ and $b$ respectively.
We take the ground state $|\chi\rangle$ to be a tensor product of
the now known
{\em one-dimensional}
ground states for the one-dimensional subalgebras $H_+$ generated
by the $a$ and $b$ oscillators (the representation of ${H^2}_+$ trivially
decomposes into a sum of such representations).
If we determine the matrix elements
involving odd numbers of both $a$ and $b$ oscillators, then
the matrix elements of the representation will be completely determined.
In fact, clearly all that we have to do is to determine
$\langle P|a_{-m}b_{-n}|0\rangle$ for all $m,n$ (since any pair
$U(a_{-m_1}b_{-n_1}|0\rangle)_0U(a_{-m_2}b_{-n_2}|0\rangle)_0$
in \reg{dooodah} may be replaced by $U_0(a_{-m_1}b_{-n_1}|0\rangle*
a_{-m_2}b_{-n_2}|0\rangle)_0$, which is even in both the $a$ and $b$
oscillators, and can be re-expressed in terms of $*$ products of states
in the two copies of $H_+$).

Trivially we see that any quasi-primary state involving terms with one
$a$ oscillator and one $b$ oscillator must be of the form
\begin{equation}
\label{dod}
\alpha(\sum_p A_p a_{-p}b_{-N+p}|0\rangle\,,
\end{equation}
(plus of course terms of other forms),
where
\begin{equation}
\label{Pi}
pA_p+(N+1-p)A_{p-1}=0\,.
\end{equation}
Thus, from \reg{solution},
we see that $\langle P|a_{-m}b_{-n}|0\rangle$ is determined in terms
of $\langle P|a_{-1}b_{-N}|0\rangle$, $N<m+n$.
We set $\langle P|a_{-1}b_{-N}|0\rangle=\rho_N$.

Now
\begin{equation}
\label{doubleU}
\langle\chi|[U(a_{-1}a_{-1}|0\rangle)_0,U(a_{-1}b_{-N})_0]|\chi\rangle
=0\,,
\end{equation}
since we have chosen $U(a_{-1}a_{-1}|0\rangle)_0|\chi\rangle=
2\gamma^a|\chi\rangle$ for some scalar $\gamma^a$,
corresponding to the conformal weight of the ground state
with respect to the copy $H^a_+$ of $H_+$ corresponding to $a$.
But from \reg{duality},
we find
\begin{equation}
[U(a_{-1}a_{-1}|0\rangle)_0,U(a_{-1}b_{-N})_0]=U(a_{-1}b_{-N}|0\rangle)_0
+U(a_{-2}b_{-N}|0\rangle)_0\,.
\end{equation}
Then \reg{doubleU}
gives
\begin{equation}
\langle\chi|U(a_{-2}b_{-N}|0\rangle)_0|\chi\rangle=-\rho_N\,.
\label{bigomega}
\end{equation}
We see from this, together with \reg{solution}, that all of the $\rho_N$
are determined in terms of $\rho_1$.

For example, the relevant term in the expansion \reg{solution} of $|P\rangle$
giving $a_{-1}b_{-2}|0\rangle$ is given by
\begin{equation}
\beta(2,1)\rho_1L_{-1}a_{-1}b_{-1}|0\rangle=
-{1\over 2}\rho_1(a_{-1}b_{-2}+a_{-2}b_{-1})|0\rangle\,,
\end{equation}
and hence $\rho_2=-\rho_1$.

At the next level we pick up a new quasi-primary state as in \reg{dod},
and so for some $\alpha\in\re$ we find the two-oscillator odd $b$ state
in $|P\rangle$ to be
\begin{eqnarray}
\beta(2,2)\rho_1{L_{-1}}^2a_{-1}b_{-1}|0\rangle&+&
\alpha(a_{-1}b_{-3}-{3\over 2}a_{-2}b_{-2}+a_{-3}b_{-1})|0\rangle=
\nonumber\\
{3\over 10}\rho_1(a_{-1}b_{-3}+a_{-2}b_{-2}+a_{-3}b_{-1})|0\rangle
&+&\alpha(a_{-1}b_{-3}-{3\over 2}a_{-2}b_{-2}+a_{-3}b_{-1})|0\rangle\,.
\end{eqnarray}
Now, \reg{bigomega} for $N=2$ gives
\begin{equation}
4\left({3\over 10}\rho_1-{3\over2}\alpha\right)=-\rho_2\,.
\end{equation}
Hence
\begin{equation}
\rho_3\equiv3\alpha+{9\over10}\rho_1=\rho_1\,.
\end{equation}
Similarly for the higher order terms.

[Note that, if desired, we may obtain directly an expression
for $\rho_N$ by the following trick, which also illustrates the sort
of manipulations required to evaluate matrix elements in ${H^2}_+$
in terms of the representation of ${H_+}^2$.

Consider
\begin{equation}
\langle\chi|U(b_{-1}b_{-1}|0\rangle)_0U(a_{-1}b_{-N}|0\rangle)_0
|\chi\rangle=2\gamma^b\rho_N\,,
\end{equation}
where $\gamma^b$ is the conformal weight of the ground state with respect
to the copy $H^b_+$ of $H_+$ corresponding to $b$.

However, using \reg{product}, we may rewrite this matrix element as
\begin{eqnarray}
\langle\chi|U(
b_{-1}b_{-1}|0\rangle&*&a_{-1}b_{-N}|0\rangle)_0|\chi\rangle
=\langle\chi|U\left(a_{-1}{b_{-1}}^2b_{-N}|0\rangle\right)_0|\chi\rangle
\nonumber\\
&&+2N\langle\chi|U\left(a_{-1}b_{-N}|0\rangle+
2a_{-1}b_{-N-1}|0\rangle+a_{-1}b_{-N-2}|0\rangle\right)_0|\chi\rangle\,,
\end{eqnarray}
using the explicit form for the $*$ product given in \reg{stardef}.

But
\begin{eqnarray}
a_{-1}{b_{-1}^2}b_{-N}|0\rangle&=&b_{-1}b_{-N}|0\rangle*a_{-1}b_{-1}|0\rangle
-(-1)^{N-1}N\sum_{p=1}^{N+2}\left({N+1\atop{p-1}}\right)
a_{-1}b_{-p}|0\rangle\nonumber\\
&&\hskip-80pt-{1\over 2}N(N+1)a_{-1}b_{-N}|0\rangle
-N(N+1)a_{-1}b_{-N-1}|0\rangle
-{1\over 2}N(N+1)a_{-1}b_{-N-2}|0\rangle\,.
\end{eqnarray}
The matrix element of the first term on the right hand side is
determined in terms of the known representation for the $H_+$ corresponding
to the $b$'s and also $\rho_1$.
Together then these expressions
provide the desired solution for $\rho_{N}$ in terms of
$\rho_m$, $m<N$ (for $N\geq 4$).]

We label the possible ground states for the one-dimensional representation
of the algebra $H_+$ corresponding to $a$ as
\begin{eqnarray}
\psi_1=|0\rangle\,,\qquad\psi_2=|\lambda\rangle\,,\qquad\psi_3=a_{-1}|0\rangle
\nonumber\\
\psi_4=\rho\,,\qquad\psi_5=c_{-{1\over 2}}\rho\,,
\end{eqnarray}
as in \reg{setstates} (we have dropped the $\pm$ subscript on $\psi_2$,
as well as the various phase factors required to ensure reality of the
states, simply for ease of notation -- note also that we explicitly take
the momentum $\lambda$ to be non-zero). Similarly we have states $\psi_i$,
$1\leq i\leq 5$, corresponding to the algebra for $b$. We may take $\chi$ to
be given by one of the 15 possible inequivalent tensor products
$\psi_i\otimes\phi_j$. Let us consider the various possibilities and identify
which correspond to ground states.

We may evaluate the norm of the state $U(a_{-1}b_{-1}|0\rangle)_2|\chi\rangle$
for $\chi=\psi_3\otimes\phi_3\equiv a_{-1}b_{-1}|0\rangle$
using the known representation structure as in the above calculations, {\em
i.e.}
\begin{eqnarray}
||U(a_{-1}b_{-1}|0\rangle)_2|\chi\rangle||^2&=&
\langle\chi|U(a_{-1}b_{-1}|0\rangle)_{-2}U(a_{-1}b_{-1}|0\rangle)_2
|\chi\rangle\nonumber\\
&=&
\oint_0{dw\over 2\pi i}\oint_{|z|>|w|}{dz\over 2\pi i}
z^{-1}w^3\langle\chi|U(a_{-1}b_{-1}|0\rangle,z)U(a_{-1}b_{-1}|0\rangle,w)
|\chi\rangle\nonumber\\
&=&\oint_0{dw\over 2\pi i}\oint_{|z|>|w|}{dz\over 2\pi i}
z^{-1}w^3\sum_n\langle\chi|U(V(a_{-1}b_{-1})_{-n}a_{-1}b_{-1}|0\rangle,w)|\chi\rangle
(z-w)^{n-2}\nonumber\\
&=&\oint_0{dw\over 2\pi i}\oint_{|z|>|w|}{dz\over 2\pi i}
z^{-1}\sum_nw^{-n+1}\langle\chi|U(V(a_{-1}b_{-1})_na_{-1}b_{-1}|0\rangle)_0|\chi\rangle
(z-w)^{n-2}\,,\nonumber\\
\end{eqnarray}
using \reg{duality}.

The matrix element may now be evaluated in terms of the known representations
$\chi$ gives of ${H_+}^2$. Note that we do not need to do this explicitly,
for we may use the known results. We have
$||U(a_{-1}b_{-1}|0\rangle)_2|\chi\rangle||^2=1$, and so we deduce
that $\psi_3\otimes\phi_3$ is not consistent as a ground state.

Similarly, we may eliminate $\chi=\psi_5\otimes\phi_5$
and $\psi_3\otimes\phi_2$.

Now consider the possible ground states $\chi=\psi_i\otimes\phi_j$ with
$i=4$, $5$, $j=1$, $2$, $3$. Intuitively, these will be non-meromorphic
representations since they are twisted in one coordinate and untwisted
in the other. Let us verify this in a particular case.

We consider $\chi=\psi_4\otimes\phi_1\equiv\rho\otimes|0\rangle$, and
the matrix element
\begin{equation}
\langle\chi|U(a_{-1}b_{-1}|0\rangle,z)
U(a_{-1}b_{-1}|0\rangle,w)|\chi\rangle\,.
\end{equation}
By \reg{duality} this is
\begin{equation}
\sum_n\langle\chi|U\left(V(a_{-1}b_{-1}|0\rangle)_{-n}a_{-1}b_{-1}|0\rangle,
w\right)|\chi\rangle(z-w)^{n-2}\,.
\end{equation}
Now, because of the vacuum in the $a$ sector, this sector is trivial, and the
only terms in the sum which contribute are
\begin{equation}
(z-w)^{-4}+\sum_{n\geq 0}\langle\rho|U(b_{-1}b_{-n-1}|0\rangle,w)|\rho\rangle
(z-w)^{n-2}\,.
\end{equation}
We then see from \reg{mystery} and \reg{delooo} that this can be evaluated
as
\begin{equation}
(z-w)^{-4}-{1\over 2}\sum_{n=0}^\infty\left({-{1\over 2}\atop n+1}\right)
{n+1\over n+2}w^{-n-2}(z-w)^{n-2}={1\over 2}{z+w\over\sqrt{zw}}(z-w)^{-4}\,,
\end{equation}
as required.

We are left with possible inequivalent ground states $\chi=a_{-1}|0\rangle$,
$|\lambda\rangle$, $\rho$ and $c_{-{1\over 2}}\rho$
($\lambda$ is now two-dimensional, and possibly zero).
As we have shown, the matrix elements of the corresponding representation of
${H^2}_+$ are determined in terms of the parameter $\rho_1=
\langle\chi|U(a_{-1}b_{-1}|0\rangle)_0|\chi\rangle$.

Now, as above, we may evaluate $||U(a_{-1}b_{-1}|0\rangle)_0|\chi\rangle||^2$.
For $\chi=\rho$, this vanishes (note again that no calculation is necessary
as we know that this is determined in terms of the known results),
and so we have $\rho_1=0$ in this case and the representation is fixed
uniquely.

For the other cases, we can in principle evaluate matrix elements of the
form
\begin{equation}
\langle\chi|U(a_{-1}b_{-1}|0\rangle)_0U(\psi)_0U(a_{-1}b_{-1}|0\rangle)_0
|\chi\rangle
\end{equation}
for $\psi\in{H_+}^2$. We will find that $U(a_{-1}b_{-1}|0\rangle)_0
|\chi\rangle$ is a ground state for a representation of ${H_+}^2$
distinct from that corresponding to
$\chi$ in the
cases $\chi=a_{-1}|0\rangle$ and $\chi=c_{-{1\over 2}}\rho$. (For example,
$V(a_{-1}b_{-1}|0\rangle)_0a_{-1}|0\rangle=b_{-1}|0\rangle$.) We must therefore
have $\langle\chi|U(a_{-1}b_{-1}|0\rangle)_0|\chi\rangle=0$,
{\em i.e.} $\rho_1=0$.

The only possible ground state left to consider is
$|\chi\rangle=|\lambda\rangle$.
An explicit calculation must show that $U(a_{-1}b_{-1}|0\rangle)_0|\chi\rangle$
is a ground state for an isomorphic representation if $\gamma^a\gamma^b\neq 0$
(in fact
$V(a_{-1}b_{-1}|0\rangle)_0|\lambda\rangle=\lambda^a\lambda^b
|\lambda\rangle$).

We will obtain these same matrix elements if we take our representation to be
given by
\begin{equation}
V\left(\prod_{i=1}^Ma_{-m_i}\prod_{j=1}^Nb_{-n_j}|0\rangle,z\right)
=:\prod_{i=1}^M{i\over (m_i-1)!}{d^{m_i}\over dz^{m_i}}X^a(z)
\prod_{j=1}^N{i\over (n_j-1)!}{d^{n_j}\over dz^{n_j}}X^b(z):\Omega^N\,,
\end{equation}
acting on a degenerate ground state consisting of copies $|\lambda\rangle_i$
of $|\lambda\rangle$,
where $X^a$ and $X^b$ are the string fields \reg{stringfield} for $a$ and
$b$ respectively and $\Omega^2=1$.

$\Omega$ is a hermitian matrix, since ${\langle\lambda|_iU(a_{-1}b_{-1}
|0\rangle)_0|\lambda\rangle_j}^\ast=
\langle\lambda|_jU(a_{-1}b_{-1}|0\rangle)_0|\lambda\rangle_i$ by the
hermitian structure of the representation.
Thus, diagonalising $\Omega$ gives the usual irreducible representations (the
arbitrary sign $\pm 1$ corresponding to the equivalence of representations
given by the map $\theta_b$ ($\theta_ba_n{\theta_b}^{-1}=a_n$,
$\theta_bb_n{\theta_b}^{-1}=-b_n$,
$\theta_ba_n{\theta_b}^{-1}=a_n$,
$\theta_ab_n{\theta_a}^{-1}=b_n$)).

The extension to more than two dimensions is similar. We start with
a state $\chi$ which is a ground state for a representation of ${H_+}^d$.
The matrix elements for arbitrary states in ${H^d}_+$ are then give as
in \reg{dooodah}.

We can argue, as before, that any matrix element
involving $U(a^i_{-m}a^j_{-n}|0\rangle)_0$ ($i\neq j$) can be reduced to ones
involving just $U(a^i_{-1}a^j_{-1}|0\rangle)_0$. We can then go through
arguments exactly as above to restrict the possible ground states $\chi$ by
rejecting ones that lead to non-meromorphic matrix elements or states not
annihilated by positive modes of the vertex operators defining
the representation. The only problem we encounter in attempting to define
the matrix elements for the representation completely is that we may
have to evaluate matrix elements containing two or more such cross terms,
{\em e.g.}
\begin{equation}
\langle\chi|\cdots U(a_{-1}^ia_{-1}^j|0\rangle)_0
U(a_{-1}^ka_{-1}^l|0\rangle)_0|\chi\rangle\,,
\end{equation}
with $i$, $j$, $k$, $l$ distinct. However, it is easy to verify (using only
the results we know,
since the calculation depends again only on the
${H_+}^d$ structure) that $||U(a_{-1}^ia_{-1}^j|0\rangle)_0
U(a_{-1}^ka_{-1}^l|0\rangle)_0|\chi\rangle||^2=0$ for all the allowable
$\chi$ (and so we can evaluate everything in terms of scalars
\begin{equation}
\rho^{ij}\equiv\langle\chi|U(a_{-1}^ia_{-1}^j|0\rangle)_0|\chi\rangle
\end{equation}
evaluated as above), except
for $|\chi\rangle=|\lambda\rangle$ (and $\lambda$ non-zero
in all four relevant coordinates). We then have the situation as above, with
the actions of the zero modes in this case given by matrices $\Omega_{ij}$
which commute for distinct indices (since the corresponding operators
$U_0$ trivially commute from \reg{duality}). Simultaneously diagonalising them
again gives the required result up to a trivial equivalence.

Finally, we must consider representations of $\Hil(\Lambda)_+$, where
$\Lambda$ is a non-zero even Euclidean lattice of dimension $d>1$.
The inclusion of momenta is essentially a straightforward extension
of the above. Before so proceeding though, we must first argue that
we can evaluate the coefficients of all terms in $|P\rangle$ given by
the action of states in $H_+^a$ and $H_+^b$ (equivalently all zero momentum
states with $\theta_a=\theta_b=1$)
on
momentum states $|\lambda\rangle_+$, $\lambda\in 2\Lambda$.
Note that
again we restrict our considerations to two dimensions for simplicity
of notation.

Suppose the representation is chosen so that $L_0^a$ and $L_0^b$ (using
the obvious notation) act as scalars $\gamma^a$ and $\gamma^b$
respectively. As in \reg{howdy} and \reg{howdydoody}, we see that for
states $\psi_a$ and $\psi_b$ in $H_+^a$ and $H_+^b$ respectively
\begin{equation}
O(\psi_a,\psi_b)=(L^a_{-1}+L_0^a)\psi_a*\psi_b\,.
\end{equation}
Now $\psi_a*\psi_b$ is trivially the state $\psi_a\otimes\psi_b$ in the
full conformal field theory. We then deduce from \reg{whodoyoudo} that we
have, for terms in $|P\rangle$ given by the action of zero momentum
states with $\theta_a=\theta_b=1$ on $|\lambda\rangle_+$, $\lambda\in
\Lambda$,
\begin{equation}
(L_1^a+L_0^a)|P\rangle=(L_1^b+L_0^b)|P\rangle=0\,.
\end{equation}
We then proceed exactly as before, and solve the separate one-dimensional
problems for $a$ and $b$ (using $L^a$ and $L^b$ in place of $L$ as
appropriate) using our previous results (note that we need here the
one-dimensional result with non-zero momentum). Finally, we use the
orthogonality
relation \reg{orthog} with $a=|\lambda\rangle_+$, $\lambda\in\Lambda$. This
clearly relates, as before, states in $|P\rangle$ with momentum $(n+1)\lambda$
to those with momentum $(n-1)\lambda$, and we are able to find all terms
in $|P\rangle$ given by the action of zero momentum states with
$\theta_a=\theta_b=1$
on momentum states $|\lambda\rangle_+$, $\lambda\in 2\Lambda$, in terms of the
representations of $H_+^a$ and $H_+^b$. (In fact, the terms for other momenta
are given
in terms of the coefficient for states $|\lambda\rangle_+$ for $\lambda$ taking
values
only in the cosets $\Lambda/2\Lambda$ -- the similarity to the gamma matrix
representations discussed in \cite{DGMtwisted} is not accidental, as will soon
become apparent).

Now, exactly as argued before in the zero momentum case,
all the matrix elements for the representation of
$\Hil(\Lambda)_+$ are known once we have the matrix elements
\begin{eqnarray}
\label{mata}
\langle|\chi|U(\psi_1)_0\cdots U(\psi_M)_0&&U(\phi_1)_0\cdots
U(\phi_N)_0|\chi\rangle\\
\label{matb}
\langle|\chi|U(\psi_1)_0\cdots U(\psi_M)_0&&U(\phi_1)_0\cdots U(\phi_N)_0
U(a_{-1}b_{-1}|0\rangle)_0|\chi\rangle\\
\label{matc}
\langle|\chi|U(\psi_1)_0\cdots U(\psi_M)_0&&U(\phi_1)_0\cdots U(\phi_N)_0
U(|\lambda\rangle_+)_0|\chi\rangle\\
\label{matd}
\langle|\chi|U(\psi_1)_0\cdots U(\psi_M)_0&&U(\phi_1)_0\cdots U(\phi_N)_0
U(a_{-1}b_{-1}|0\rangle)_0U(|\lambda\rangle_+)_0|\chi\rangle\,,
\end{eqnarray}
where $\psi_i\in H_+^a$, $\phi_j\in H_+^b$, $\lambda\in\Lambda$.

We start with $\chi$ a product of representations for  $H_+^a$ and $H_+^b$
as before, and follow essentially the same argument. The possible ground states
are as we found before. [Note that we can now say slightly more in the case of
a representation with ground state $|\mu\rangle$. We find
$||U(|\lambda\rangle_+)_n
|\mu\rangle||^2\neq0$ for some $n>0$ if $\mu$ is not of minimal norm in the
set $\mu+\Lambda$ (we may calculate this norm using the structure of
$|P\rangle$ which we know so far). That we also require $\mu\in
\Lambda^\ast$ is easy to deduce by evaluating the four-point
function as in \reg{4pfn}.]

It is now simply a question of considering the separate possibilities
as we did before. Since \reg{mata} and \reg{matb} are known, we need only
evaluate \reg{matc} and \reg{matd}.

First consider the ground state $a_{-1}|0\rangle$. Note that
$||U(|\lambda\rangle_+)_0a_{-1}|0\rangle||^2=0$, except for $\lambda^2=2$
(we know this from the expected structure, and have enough of $|P\rangle$
to calculate it uniquely -- hence again no calculation is necessary).
For $\lambda^2=2$, we know that we may pick up (for $\lambda^a\neq 0$)
the weight one state $|\lambda\rangle_+$. We can see this at our level of
knowledge here by calculating matrix elements of the form
\begin{equation}
\langle\chi|U(|\lambda\rangle_+)_0U(\psi_1)_0\cdots U(\psi_M)_0
U(\phi_1)_0\cdots U(\phi_N)_0U(|\lambda\rangle_+)_0|\chi\rangle\,.
\end{equation}
This will show that $U(|\lambda\rangle_+)_0a_{-1}|0\rangle$ gives a
representation
of $H_+^a$, $H_+^b$ distinct from that corresponding to $a_{-1}|0\rangle$, and
we deduce that, in this case, \reg{matc} vanishes. Similarly, \reg{matd} also
vanishes.

The argument for a ground state $|\mu\rangle$ is also straightforward. We know
that $V(|\lambda\rangle_+)_0|\mu\rangle_+$ contains terms of momentum
$\pm\mu\pm
\lambda$. Thus again $U(|\lambda\rangle_+)_0|\mu\rangle_+$
can be seen to give rise to
representations of $H_+^a$ and $H_+^b$ distinct from the original,
except in the case $\lambda=\pm 2\mu$. Thus, we can argue that,
for $\lambda\neq\pm 2\mu$, \reg{matc} and \reg{matd} vanish. The
undetermined scalars $\langle\chi|U(|2\mu\rangle_+)_0|\chi\rangle$
and $\langle\chi|U(a_{-1}b_{-1}|0\rangle)_0
U(|2\mu\rangle_+)_0|\chi\rangle$ can
be easily calculated by requiring the representation to be consistent
({\em i.e.} requiring the operator product expansion
of $U(|2\mu\rangle_+,z)$ with itself to be what it should
be from \reg{duality}), or we
can simply restrict to the case of a self-dual lattice $\Lambda^\ast
=\Lambda$ (in which case $\mu\in\Lambda$ and $\lambda\in 2\Lambda$, and
\reg{orthog} with $a=|\lambda\rangle_+$ fixes the scalars
in terms of the known zero-momentum states in $|P\rangle$).

For a twisted ground state $\rho$, $U(a_{-1}b_{-1}|0\rangle)_0
U(|\lambda\rangle_+)_0\rho=0$ (as we can see by evaluating the norm),
but $U(|\lambda\rangle_+)_0\rho$ gives us a ground state of a representation
which turns out to be identical to that generated from $\rho$.
We thus find vertex operators as in \reg{twistedvops}, but with some
matrix, say $M_\lambda$, in place of the gamma matrix $\gamma_\lambda$.
Note that, by the comments at the end of the paragraph preceding \reg{mata},
$M_\lambda$ is only arbitrary for one state from each coset $\Lambda/2
\Lambda$.  Locality of the vertex operators (the analog of
\reg{locality} for the $U$'s -- see \cite{DGMtriality}) then implies
that
\begin{equation}
M_\lambda M_\mu=(-1)^{\lambda\cdot\mu}M_\mu M_\lambda\,,
\end{equation}
and we recover the usual gamma matrices.

Finally, we consider a ground state $c_{-{1\over 2}}\rho$. We know
that
\begin{equation}
V_T(|\lambda\rangle_+)_0c_{-{1\over 2}}\hat\rho=2^{-\lambda^2}
\left((1-2{\lambda^1}^2)c_{-{1\over 2}}-2\lambda^1\lambda^2
d_{-{1\over 2}}\right)\gamma_\lambda\hat\rho\,,
\end{equation}
for some spinor ground state $\hat\rho$. Then
\begin{equation}
R(\lambda)_0c_{-{1\over 2}}\hat\rho\equiv
2^{-3}\left(3^22^{\lambda^2}U(|\lambda\rangle_+)_0
-2^{9\lambda^2}U(|3\lambda\rangle_+)_0\right)c_{-{1\over 2}}\hat\rho
=c_{-{1\over 2}}\hat\rho\,,
\end{equation}
and similarly
\begin{equation}
U(a_{-1}b_{-1}|0\rangle)_0R(\lambda)_0c_{-{1\over 2}}\hat\rho
=d_{-{1\over 2}}\hat\rho\,,
\end{equation}
where $d$ is the other twisted oscillator. Thus, using this combination
in place of $U(|\lambda\rangle_+)_0$ (and remembering the comment at the end
of the paragraph preceding \reg{mata}), we find that
the modified \reg{matd} vanishes
and \reg{matc} gives rises to again what turns out to be the gamma matrices.

This completes the sketch of the argument that the representations
of $\Hil(\Lambda)_+$ comprise only the known untwisted and $\ze_2$-twisted
representations detailed earlier in this paper.

Let us conclude this section with a simple observation.
For a representation of $\Hil(\Lambda)_+$ with ground
state $|\mu\rangle_\pm$,
$\mu\in\Lambda^\ast$, the corresponding $P$ has terms in
$|2\mu\rangle_+$.
However, these can only contribute to matrix elements if $2\mu\in\Lambda$,
and further we see that we must have $2\mu^2$ even (so that this is a
quasi-primary state of even weight in $P$) -- in other words, matrix elements
with the corresponding terms in $P$ must vanish otherwise, and we can
consistently set them to zero, as required by the equations satisfied by $P$.
Thus note that if all allowed representations of $\Hil(\Lambda)_+$ have all
terms in $P$ not excluded from contributing to matrix elements by this
argument, this is the same as saying $\sqrt 2\Lambda^\ast$
is even, {\em i.e.} which is the same as saying that the $\ze_2$-orbifold
$\Hil(\Lambda)_+\oplus\Hil_T(\Lambda)_+$
is consistent \cite{DGMtwisted,thesis}.
Further pursuit of such a point of view may enable us to better understand
this condition on the lattice. (Note that consistency of the orbifold
theory in this notation is decided by consideration of matrix elements
of the form
\begin{equation}
\langle P|V(\psi_1,z_1)\cdots V(\psi_N,z_N)|P\rangle\,.)
\end{equation}
\section{Conclusions}
\label{conclusions}
We have proven that the known representations of the
reflection-twisted Heisenberg algebra
comprise the complete set of modules, and shown that the same
is true in first one dimension and then given an argument
that it is true in
general for the reflection-twisted FKS lattice conformal field
theories.
In particular, we have found an alternative derivation
for the rather mysterious term $\Delta(z)$
which occurs in the twisted sector vertex operators \reg{mystery}, initially
found in \cite{DGMtwisted,thesis} by a rather ad hoc correction of a normal
ordering
problem. We find it in the solutions $P_b$ and $P_c$ to our constraint
equations \reg{solution}, \reg{L2reln} and \reg{orthog}, though we would like
to understand more clearly the relation of the form of $\Delta$ to the
embedding
of $O(H_+)$ in $O(H)$ from which the twisted representation solutions arise.
Also, we wish to understand better the relation of our work to that
of Zhu, as well as the work of Dong, Li and Mason \cite{DLMZhu} who generalize
Zhu's theory to twisted modules.

It must be stressed that, though the uniqueness of the $\ze_2$-twisted
representation is the main result of this paper, the method used is of
significance and is applicable to many other cases.
In particular, there is an obvious extension to higher order twisted
modules \cite{PSMthird,ZpOrbifold}. In this paper, we have used the known
results for the structure of the twisted modules in several places.
It would be interesting to see also what happens in cases other
than the simple twisted cases where we have less prior knowledge
of the representation theory.

Also one can attempt to use the techniques employed here to develop a
systematic means of classifying representations
of any given meromorphic conformal field theory.
In addition, one may try to construct orbifolds of theories for which there
is no obvious
geometric interpretation in terms of the propagation of a string on
a singular manifold, since our point of view provides abstract tools
not constrained by any
requirement for an explicit construction (see the comments at the end
of \cite{PSMreps}).
The analysis of
the construction of orbifolds, and in particular the concept of an inverse
to such a construction, by such techniques may help to realize the
ideals of treating the original conformal field theory and its orbifold
on the same footing \cite{PSMSchell,MichaelHep}

Our method, in much the same way as that of Dong in the case of the
Leech lattice \cite{dongmoonrep}, is untidy however in comparison with
that of \cite{PSMreps} (though it does have the advantage
of revealing explicitly how the twisted structure arises).
Though the argument of
\cite{PSMreps}, as discussed in the
introduction, can only be regarded as heuristic at present, it would
therefore appear judicious to attempt to tighten it. The main
problem was the potentially non-analytic behavior of the matrix elements
for a representation of the FKS theory
defined in terms of some $P$ satisfying the constraint equations.
However, because matrix elements involving pairs of vertex operators
for states in the odd-parity sector of the FKS theory can be rewritten
in terms of those of the reflection-twisted projection, we can really
only expect at worst square root branch cuts in any correlation function.
We would expect similar results for any extension of a representation
of a conformal field theory to a larger one in which it is
{\em finitely} embedded.
The formalizing of this rough argument is work in progress. Similar ideas
occur in \cite{DongInduced}, and further investigation is required to
elucidate the links. Note though that in general the techniques developed
in the present
paper will be required to analyze the representations
of an arbitrary meromorphic conformal field theory, {\em i.e.}
when no embedding of the conformal field theory in a larger simpler
theory is available.
\section{Acknowledgments}
The author would like to thank the Royal Society for funding this work
via a Commonwealth Fellowship and the University of Adelaide for its
hospitality during the tenure of this Fellowship.

In addition, he would also like to thank Michael Tuite for bringing the work
of Zhu to his attention and for many interesting exchanges of views on that and
related topics. Thanks also to Jim McCarthy for useful
discussions, particularly on the work of section \ref{notsosimple},
and to Klaus Lucke for numerous corrections.

\end{document}